\documentclass{pasj01}
\usepackage{color}
\usepackage{lineno}

\Received{}
\Accepted{}
 
\begin{document} 
\title{Physics of nova outbursts: Theoretical models of classical nova 
outbursts with optically thick winds on $1.2~M_\odot$ and $1.3~M_\odot$ 
white dwarfs
}

\author{Mariko \textsc{Kato}\altaffilmark{1}}
\altaffiltext{1}{Department of Astronomy, Keio University, Hiyoshi, Yokohama
  223-8521, Japan}
\email{mariko.kato@hc.st.keio.ac.jp}

\author{Hideyuki \textsc{Saio},\altaffilmark{2}}
\altaffiltext{2}{Astronomical Institute, Graduate School of Science,
    Tohoku University, Sendai, 980-8578, Japan}

\author{Izumi \textsc{Hachisu}\altaffilmark{3}}
\altaffiltext{3}{Department of Earth Science and Astronomy, College of Arts and
Sciences, The University of Tokyo, 3-8-1 Komaba, Meguro-ku, Tokyo 153-8902, Japan}

\maketitle
\begin{abstract}
We present time-dependent nova outburst models with optically thick winds 
for a 1.2 and 1.35 $M_\odot$ white dwarfs (WDs) with a mass accretion rate of
$5 \times 10^{-9}~M_\odot$~yr$^{-1}$ and for a 1.3 $M_\odot$ WD with
$2 \times 10^{-9}~M_\odot$~yr$^{-1}$.
The X-ray flash occurs 11 days before the optical peak of the
1.2 $M_\odot$ WD and 2.5 days before the peak of the 1.3 $M_\odot$ WD. 
The wind mass loss rate of the 1.2 $M_\odot$ WD (1.3 $M_\odot$ WD)
reaches a peak of $6.4 \times 10^{-5}~M_\odot$ yr$^{-1}$
($7.4 \times 10^{-5}~M_\odot$ yr$^{-1}$) at the epoch of the maximum
photospheric expansion with the lowest 
photospheric temperature of $\log T_{\rm ph}$ (K)=4.33 (4.35). 
The nuclear energy generated during the outburst is lost in a form of 
radiation (61\% for the 1.2 $M_\odot$ WD; 47\% for the 1.3 $M_\odot$ WD), 
gravitational energy of ejecta
(39\%; 52\%), and kinetic energy of the wind (0.28\%; 0.29\%).
We found an empirical relation for fast novae 
between the time to optical maximum from the outburst 
$t_{\rm peak}$ and the expansion timescale $\tau_{\rm exp}$.
With this relation,
we are able to predict the time to optical maximum $t_{\rm peak}$
from the ignition model (at $t=0$) without following a time-consuming
nova wind evolution.
\end{abstract}


\section{Introduction}\label{introduction}


A nova is a thermonuclear runaway event on a mass-accreting white dwarf (WD) 
\citep{pri95, kov98, yar05,epels07, she09, den13hb, ida13, wol13, kat14shn,
kat15sh, tan14,taji15,chen19,jos20}.
Nova winds are accelerated owing to radiation pressure-gradient 
deep inside the photosphere \citep{fri66, fin71, zyt72, rug79}. 
Stellar evolution codes, however, meet numerical difficulties 
in calculation beyond the extended stages of nova outbursts 
(\citet{kat17palermo} for a review). 
\citet{kat17} showed, for the first time,
a way how to calculate the whole cycle
of nova outburst including radiative acceleration in a self-consistent way.
Based on their results, \citet{kat22sh} presented 
the classical nova model for a $1.0~M_\odot$ WD with a mass accretion rate 
of $\dot M_{\rm acc}=5 \times 10^{-9}~M_\odot$~yr$^{-1}$.
This is a unique nova model that self-consistently includes
radiation-pressure-gradient acceleration in the outburst evolution
calculation.  The present work adds two more one cycle models 
of classical novae 
($1.2~M_\odot$ and $1.3~M_\odot$ WDs), which enable us to quantitively
compare the nova outbursts for different WD masses. 

This paper is organized as follows.
Section \ref{section_model} presents our models 
including X-ray light curves and internal structures. 
Section \ref{sec_energy} reports energy budget in the nova outburst for 
each of our model.
Discussion and concluding remarks follow in sections \ref{sec_discussion}
and \ref{conclusions}, respectively.



\begin{figure}
  \includegraphics[width=8cm]{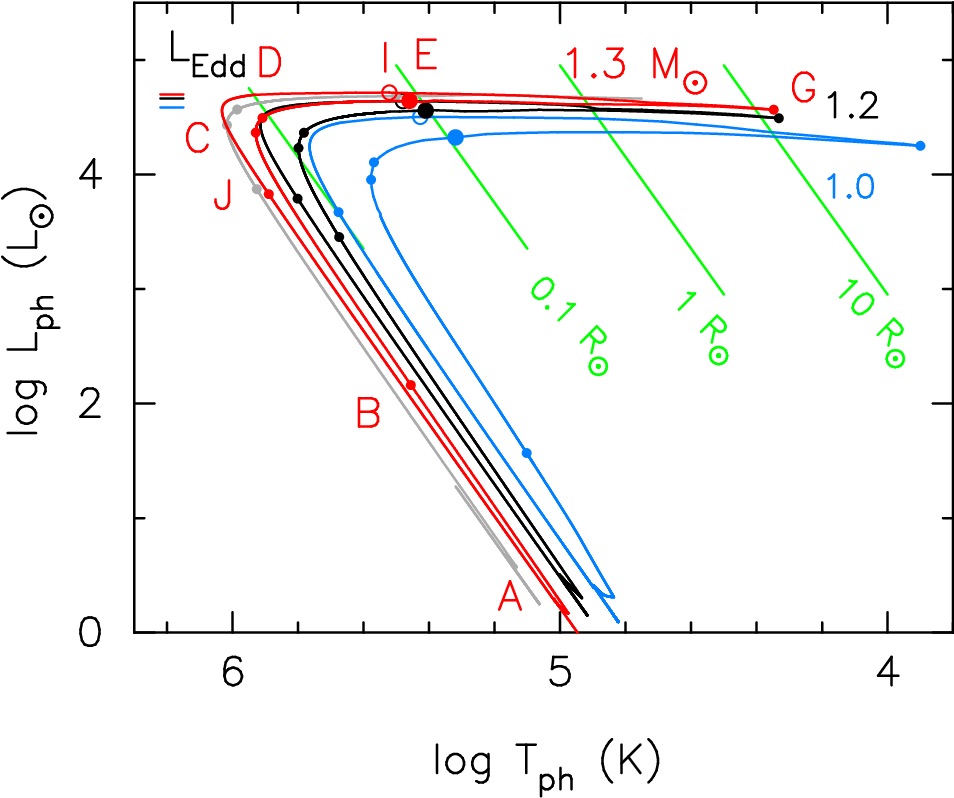}
\caption{The HR diagrams of one cycle of hydrogen shell flashes for
our outburst models of 1.0 $M_\odot$ (blue line), 1.2 $M_\odot$
(black line), 1.3 $M_\odot$ (red line), and 1.35 $M_\odot$ (light gray) 
WDs.  The 1.35 $M_\odot$ model is shown only for the rising phase (up to
before maximum expansion of the photosphere).     
Selected stages during a shell flash are denoted 
counterclockwise direction 
starting from the bottom of the cycle.  
A: Quiescent phase before the shell flash. 
B: The epoch when $L_{\rm nuc}$ reaches maximum ($t=0$).
C: Highest photospheric temperature in the rising phase. 
D: Peak of X-ray flash (0.3 keV - 1.0 keV). 
E: Wind begins to emerge from the photosphere. 
G: The maximum expansion when both the wind mass loss rate and photospheric 
radius reach maximum.
I: The wind mass-loss stops (open circle). The supersoft X-ray phase starts. 
J: The supersoft X-ray luminosity (0.3 keV- 1.0 keV) decreases 
to one tenth of the maximum value. 
The short horizontal line indicates the Eddington luminosity
(Equation (\ref{equation_Edd}))
for 1.3, 1.2, and 1.0 $M_\odot$ WD models.  
The thick straight green lines show a locus of constant photospheric
radius of $R_{\rm ph}= 10$, 1, 0.1, and $0.01 ~R_\odot$. 
\label{hr}}
\end{figure}

\section{Nova models} \label{section_model}

We have calculated nova models of 1.2 
and $1.35 ~M_\odot$ WDs accreting matter
at a rate of $\dot M_{\rm acc}=5\times10^{-9}M_\odot$ yr$^{-1}$. 
The chemical composition of accreting matter is assumed to be 
solar, $X=0.7$, $Y=0.28$, and $Z=0.02$. 
The mass accretion rate of $\dot{M}_{\rm acc}=5\times10^{-9}M_\odot$ yr$^{-1}$ 
is a typical value for classical novae (e.g., a central observed value of
$\dot{M}_{\rm acc}=3\times10^{-9}M_\odot$ yr$^{-1}$ given by \citet{sel19};
or close to the center in the $\dot{M}_{\rm acc}$ distribution of galactic
novae in Fig.6 of \citet{hac20skhs}). 
With the 1.0 $M_\odot$ model \citep{kat22sh}, we have a set of 
three WD mass (1.0, 1.2, and 1.35 $M_\odot$) for 
$\dot M_{\rm acc}=5\times10^{-9}M_\odot$ yr$^{-1}$. 
This enables us to see the dependence of nova properties  
on the WD mass, although only an early part of nova explosion is presented
for the 1.35 $M_\odot$.  We also calculate a $1.3\,M_\odot$ WD with 
$\dot M_{\rm acc}= 2\times10^{-9}M_\odot$ yr$^{-1}$.
We increased carbon mass fraction of the 
hydrogen-rich envelope by 0.1 and decreased 
helium mass fraction by the same amount, at the beginning of ignition, 
to mimic heavy element enhancements in ejecta, 
which are often observed in novae.  This $X_{\rm C}= 0.1$ model is
one of the candidate models for YZ Ret that satisfy
the observational duration of the short X-ray flash. 
Here, we have calculated a full cycle of the model. On the other hand,
only a very early phase of nova outburst (the X-ray flash phase)
is published in model EF of \citet{kat22shc}. 

We calculate nova evolution with a Henyey-type evolution code in 
which the optically thick winds are consistently included as a boundary 
condition of the evolution calculation.
We used the same computer code and calculation method as in \citet{kat22sh}; 
We stop the mass accretion when the photospheric luminosity
$L_{\rm ph}$ increases to
$\log L_{\rm ph}/L_\odot=3.5$ and resumes when the luminosity decreases
less than $\log L_{\rm ph}/L_\odot=2.5$.
We use the OPAL opacity tables \citep{igl96}.  

More exactly, we first calculate several cycles of nova evolution 
with the Henyey-type code 
with temporarily assumed mass-loss rate as a function of time. 
Then we obtain the steady wind solution \citep{kat94h} 
that fits smoothly to the interior 
structure at each step of evolution with mass loss. 
If the steady-state wind mass-loss rate deviates significantly 
from that assumed in the evolution calculation, 
we recalculate the whole evolution cycle with different mass-loss rate. 
We iterate this process until we have a steady wind solution that 
(1) fits smoothly to the inner structure 
and (2) the obtained wind mass-loss rate converges to 
the assumed value of the Henyey code. 
This iteration takes much human-time. 
In the 1.35 $M_\odot$ WD model, this method does not work well 
in the extended stage of the outburst.

The model parameters are summarized in Table \ref{table_models}; 
the WD mass, mass accretion rate in quiescent phase, the carbon enhancement 
(C-mix), central temperature of the WD ($T_{\rm WD}$), recurrence period, 
mass accreted until the thermonuclear runaway sets in ($M_{\rm acc}$), 
ignition mass ($M_{\rm ig}$), maximum temperature
in the hydrogen nuclear burning layer through an outburst period,
and maximum nuclear luminosity. 
The accreted mass is smaller than 
the ignition mass
because there was leftover hydrogen-rich layer when the previous outburst 
had finished.

Among three models (1.0 $M_\odot$, 1.2 $M_\odot$ and 1.35 $M_\odot$ WDs) 
with $\dot M_{\rm acc}=5\times10^{-9}M_\odot$ yr$^{-1}$, 
the 1.0 $M_\odot$ WD model is taken from \citet{kat22sh}. 
As the mass accretion rate and composition is the same, 
the main difference of these three models is the gravitational energy 
of the WD. More massive WDs ignite with smaller ignition masses because a
stronger gravity yields a larger gravitational heating at the bottom of the 
hydrogen-rich envelope.  Thus, thermonuclear runaway begins at a
smaller ignition mass than in less massive WDs. 
As a result, the recurrence period is shorter in more massive WDs.

The last column in Table \ref{table_models} shows
the maximum nuclear burning rate $L_{\rm nuc}^{\rm max}$.  
The carbon enhanced model of the 1.3 $M_\odot$ WD shows much
larger $L_{\rm nuc}^{\rm max}$ than the others because hydrogen burning
is accelerated by carbon enrichment through the CNO cycle.

\begin{longtable}{*{11}{c}}
 \caption{Characteristic properties of a shell flash model}\label{table_models}
  \hline              
    model& $M_{\rm WD}$ & $\dot M_{\rm acc}$ &C-mix&$\log T_{\rm WD}$ & $P_{\rm rec}$&$M_{\rm acc}$ &$M_{\rm ig}$ &$\log T^{\rm max}$ & $\log L_{\rm nuc}^{\rm max}$   \\ 
    & ($M_\odot$)&($M_\odot$~yr$^{-1}$)&& ({\rm K}) &(yr)&($M_\odot$) &($M_\odot$) &  ({\rm K}) & ($L_\odot$) \\
\endfirsthead
  \hline
  Name & Value1 & Value2 & Value3 \\
\endhead
  \hline
\endfoot
  \hline
\endlastfoot
  \hline
  M13C&   1.3  &$2 \times 10^{-9}$ &0.1&7.689 &1000  &$2.00\times 10^{-6}$ &$2.15 \times 10^{-6}$ &8.24  &9.65\\
  M135&   1.35  &$5 \times 10^{-9}$ &0.&7.786 &224  &$1.12\times 10^{-6}$ &$1.27 \times 10^{-6}$ &8.32 & 8.56\\
  M12&   1.2  &$5 \times 10^{-9}$ &0.&7.730 &1450  &$7.26\times 10^{-6}$ &$8.24 \times 10^{-6}$ &8.26  &8.55\\
  M10&   1.0 & $5 \times 10^{-9}$ &0. &7.687 &5370  &$2.68\times 10^{-5}$ &$3.04 \times 10^{-5}$ &8.18  &8.36\\
\end{longtable}



\subsection{H-R diagram}\label{sec_hr}

Figure \ref{hr} shows one cycle of shell flashes in the HR diagrams. 
It should be noted that the track of the 1.35 $M_\odot$ model is not
a full cycle but ended at $\log T$ (K)=4.75 
in the pre-maximum/rising phase because of numerical difficulty.
Before onset of the hydrogen shell flash, each accreting WD 
stays around the bottom of the loop (denoted by label A).
After thermonuclear runaway sets in, the photospheric luminosity
quickly increases keeping the photospheric radius almost constant.
We define the onset of the shell flash ($t=0$)
by the time when $L_{\rm nuc}$ reaches maximum. This epoch is 
denoted by label B in Figure \ref{hr}. 
After that, the nova evolves toward point C, where 
the photospheric temperature reaches maximum $T_{\rm ph}^{\rm max}$,
and then toward point D (peak of X-ray flash). When the envelope further
expands, optically thick winds begins to emerge from the photosphere
at point E, and then the matter acceleration becomes stronger.
This emergence of wind occurs at $\log T_{\rm ph}$ (K) $\sim 5.2-5.3$, 
close to the Fe opacity peak \citep{igl96}.
The photospheric radius attains its maximum and the wind
mass loss rate also reaches maximum at point G.
In the decay phase after point G, the nova track turns back to the left in
the HR diagram.  The photospheric radius decreases
and the temperature $T_{\rm ph}$ increases with time.
The winds stop at point I (open circle). 
The supersoft X-ray phase corresponds to the nova track from point I to  
point J, where the supersoft X-ray luminosity decreases down to one tenth
of its maximum value. 
The time at each epoch is summarized in Table \ref{table_time}.

The Eddington luminosity defined by
\begin{equation}
L_{{\rm Edd}} \equiv {4\pi cG{M_{\rm WD}} \over\kappa_{\rm el}}
\label{equation_Edd}
\end{equation}
is indicated with short horizontal bars in the upper-left of
Figure \ref{hr} for each WD mass, where
$\kappa_{\rm el} = 0.2 (1+X)$ g$^{-1}$ cm$^2$ is the electron scattering
opacity and $X$ is the hydrogen content.
The photospheric luminosity hardly exceeds the Eddington luminosity 
and evolves horizontally before and after the maximum expansion 
as already reported in \citet{kov98} and \citet{den13hb}.

\begin{longtable}{*{11}{l}}
 \caption{Time since the beginning of the flash}
\label{table_time}
 \hline              
      Stage & B & C & D & E & F & G & I & J \\
     & $L_{\rm nuc}^{\rm max}$ &$T_{\rm ph}^{\rm max}$ &X flash  & wind starts & T5 & max exp & wind ends
       & $0.1~L_{\rm X}^{\rm max}$\\
\endfirsthead
\endhead
  \hline
\endlastfoot
  \hline
  {\bf M13C10}\\
Time (d)      & 0.0         & 0.00073 & 0.0018  &  0.072 & 0.71 & 2.5  & 69& 96 \\
$\log T_{\rm ph}$ (K)  & 5.46 & 5.93 & 5.91 & 5.46 & 5.00  & 4.35  & 5.52 &5.89\\ 
  \hline
  {\bf M135 }\\
Time (d)         & 0.0& 0.0030  &0.0095  &0.25 &-- &7.2 & --& -- \\
$\log T_{\rm ph}$ 
(K)  & 5.93 & 6.02 &  5.99 & 5.50 &-- & 4.75 & --   &--\\ 
  \hline
  {\bf M12}\\
Time (d)         & 0.0& 0.012  &0.024  &0.77 & 2.7 & 11  & 225& 596 \\
$\log T_{\rm ph}$ 
(K)  & 5.674 & 5.80 & 5.78 & 5.41 & 5.00  & 4.33 & 5.48 &5.80\\ 
\hline
 {\bf M10 }\\
     Time (d)  & 0.0 & 0.063 & 0.11 & 1.05 & 2.83&26.0 & 530&6.46 yr \\
$\log T_{\rm ph}$~({\rm K}) &5.10 &5.58& 5.57&5.32 &5.01& 3.90 &5.43& 5.68\\
\end{longtable}

\begin{figure}
  \includegraphics[width=8cm]{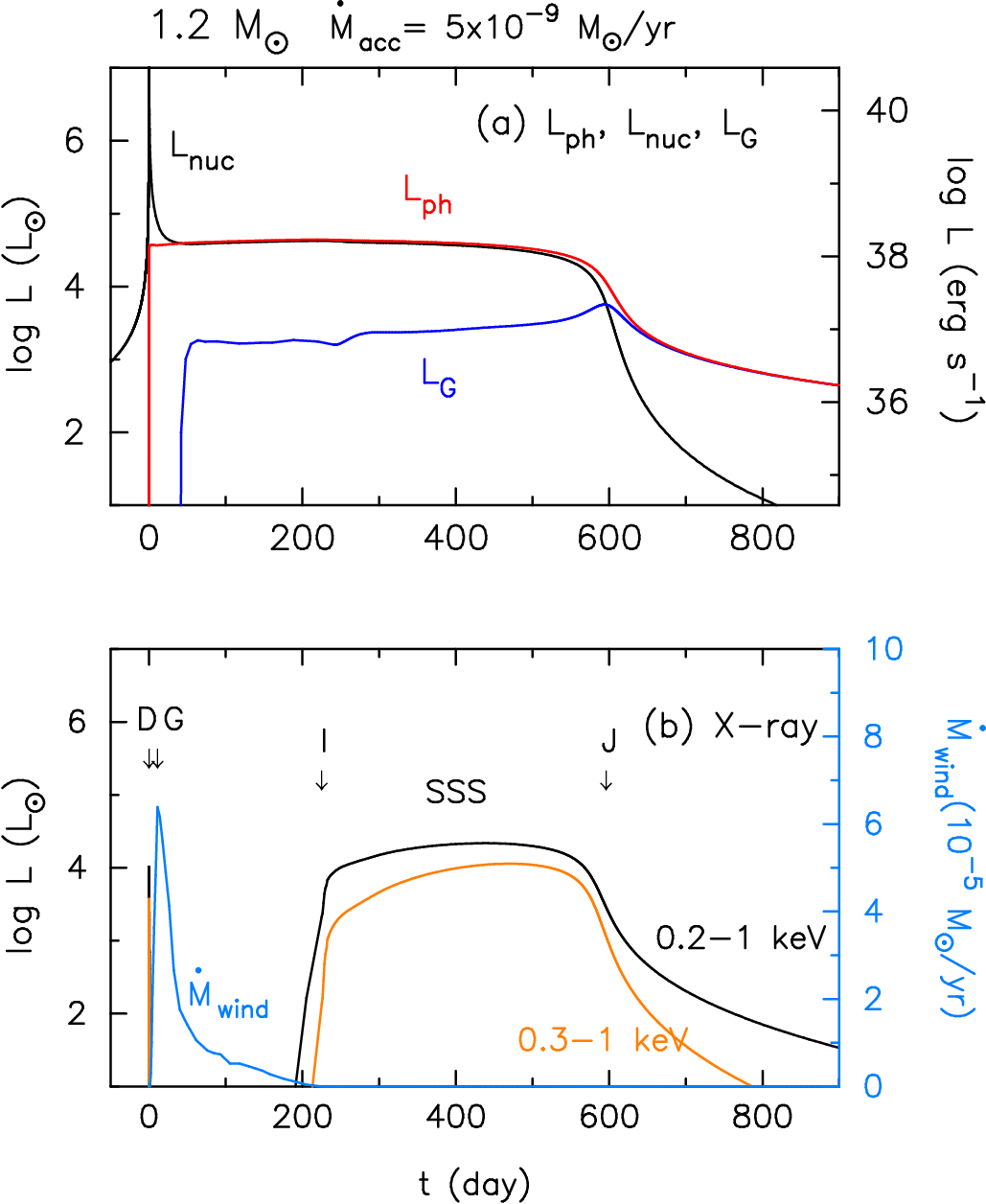}
\caption{(a) The temporal variations of the photospheric luminosity
$L_{\rm ph}$ (red line), total nuclear burning energy release rate
$L_{\rm nuc}$ (black line), and total gravitational energy release rate
$L_{\rm G}$ (blue line) for the 1.2 $M_\odot$ WD. 
We stopped the mass accretion at $t=2.6 \times 10^{-4}$ day 
and restarted at $ t=1070$ day.
(b) The supersoft X-ray light curves for the energy band of 
0.2--1.0 keV (black line) and 0.3--1.0 keV (orange line).   
Time $t=0$ corresponds to the epoch B in Fig.\ref{hr}. 
The four epochs are indicated by the four downward arrows: 
Epoch D (peak of X-ray flash), G (maximum photospheric expansion, i.e.,  
peak of wind mass-loss rate), I (end of wind phase),
and J (supersoft X-ray luminosity decreases to one tenth of its maximum). 
}\label{light.m12}
\end{figure}

\begin{figure}
\includegraphics[width=8cm]{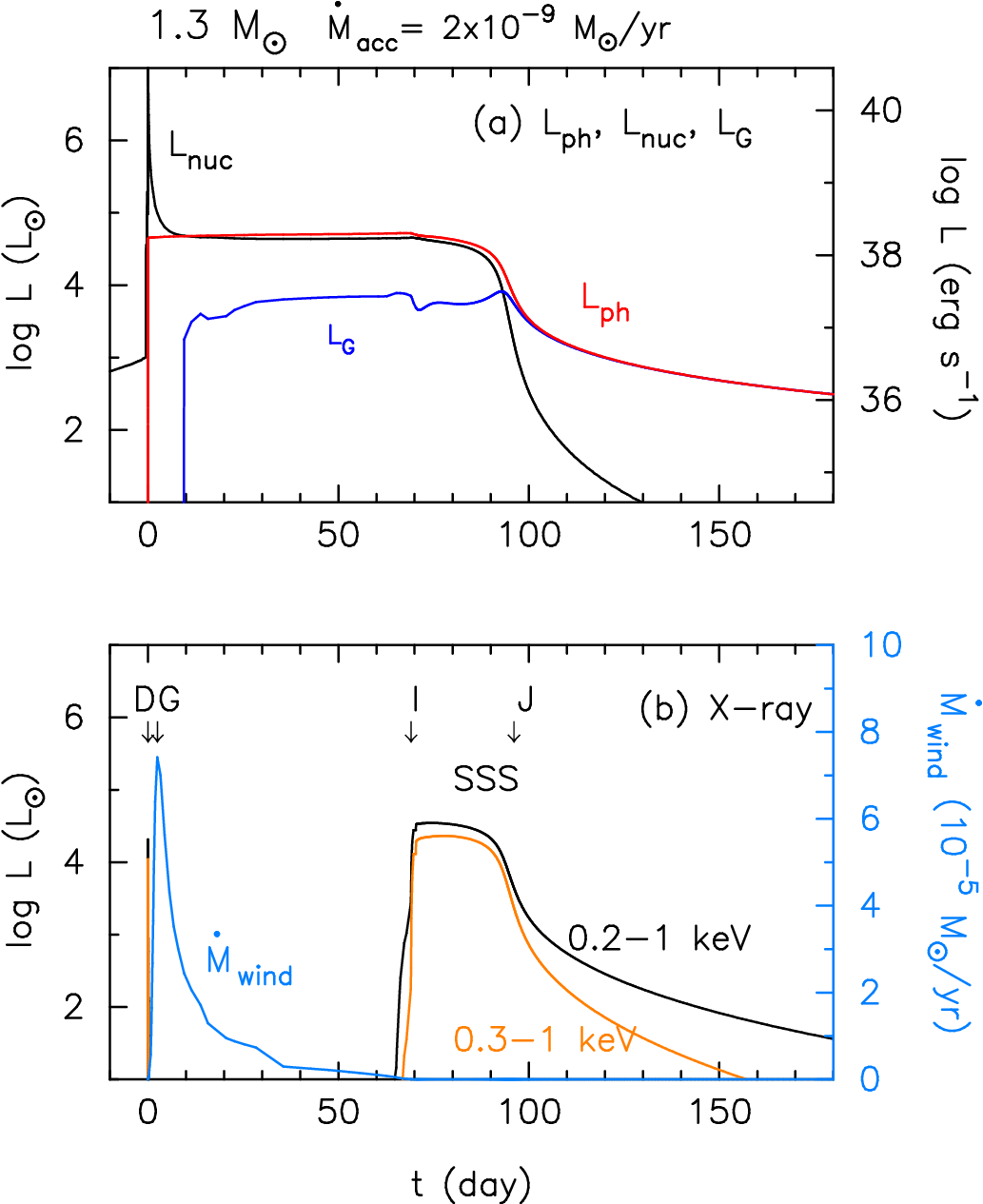}
\caption{Same as those in Fig. \ref{light.m12}, but for the 1.3 $M_\odot$
WD.  The mass accretion stopped at $t=0$ and resumed on $t=190$ day. 
}\label{light.m13}
\end{figure}


\subsection{Evolution of bolometric luminosity and X-ray light curve}

Figures \ref{light.m12}(a) and \ref{light.m13}(a)
show the temporal variations of the
emergent bolometric luminosity
$L_{\rm ph}$, total nuclear energy release rate
$L_{\rm nuc}=\int \epsilon_{\rm nuc} \delta m$,
and integrated gravitational energy release rate
$L_{\rm G}=\int \epsilon_{\rm g} \delta m$ for the 1.2 $M_\odot$ and
1.3 $M_\odot$ WDs, respectively.
Here, $\epsilon_{\rm nuc}$ and $\epsilon_{\rm g}$ are energy generation
rates per unit mass owing to nuclear burning and gravitational
energy release, respectively, and $\int \delta m$ is the integration
on the mass.

In the very beginning of the nova outburst, the nuclear energy generation
rate $L_{\rm nuc}$ amounts $\log L_{\rm nuc}/L_\odot= 9.65$ on the
1.3 $M_\odot$ WD and 8.55 on the 1.2  $M_\odot$ WD, which are much larger
than each photospheric luminosity $L_{\rm ph}$. 
A large amount of the generated energy is consumed to lift up the degeneracy, 
heat, and expand the envelope ($L_{\rm G} < 0$) in the nuclear burning
region.  As a result, the net energy flux at the surface is less than
the Eddington luminosity. 
The thermal energy is stored as the gravitational energy ($L_{\rm G}<0$)
in the very early phase, but it is emitted later with a low luminosity, 
$0 < L_{\rm G} \lesssim 0.1 ~L_{\rm ph}$, as shown in Figures
\ref{light.m12}(a) and \ref{light.m13}(a).
Thus, the photospheric luminosity does not much exceeds the 
Eddington luminosity ($L_{\rm ph} \lesssim L_{\rm Edd}$). 

Note that the super-Eddington luminosity is often observed in various
nova outbursts (e.g., \citet{del20i}), 
which is explained as the contribution of 
free-free emission from optically-thin plasma just outside the photosphere 
of a nova \citep{hac06kb, hac15k}.  However,
this problem is beyond the scope of the present work. 
(See Figure 9 and equation (B3) of \citet{hac23yzret} for the 1.0
$M_\odot$ WD.)


Figures \ref{light.m12}(b) and \ref{light.m13}(b) show 
the X-ray light curves calculated for the two energy bands of 0.2-1.0 keV
and 0.3-1.0 keV, for the 1.2 $M_\odot$ and 1.3 $M_\odot$ WDs, 
respectively. 
Both the total nova duration (H-burning on) and the SSS phase
are much shorter on the 1.3 $M_\odot$ WD. 
The X-ray turn-on time, X-ray turnoff time,
and the nova duration have been used 
to estimate the WD mass \citep{hac10k, wol13, kat20, hac24v339del}.

The X-ray flux in figures \ref{light.m12}(b) and \ref{light.m13}(b) 
show two peaks. 
The earlier X-ray peak (X-ray flash) is slightly fainter than 
the later SSS phase, which is common among all the shell flash models, 
because the maximum temperature and luminosity in the X-ray flash phase
are lower than those in the SSS phase (see Fig. \ref{hr}). 
The duration of the SSS phase is much longer than that
in the X-ray flash phase. 
In the SSS phase, the nova evolves in a thermal timescale 
$L_{\rm ph} \sim L_{\rm nuc}$. On the other hand, 
the nuclear luminosity is much larger than the photospheric luminosity
($L_{\rm nuc} \gg L_{\rm ph}$) in the X-ray flash phase. 
The difference, $L_{\rm nuc} - L_{\rm ph}$, is mainly consumed to 
expand the envelope matter quickly, which terminates 
the high temperature stage in a short time.

\begin{figure}
  \includegraphics[width=8cm]{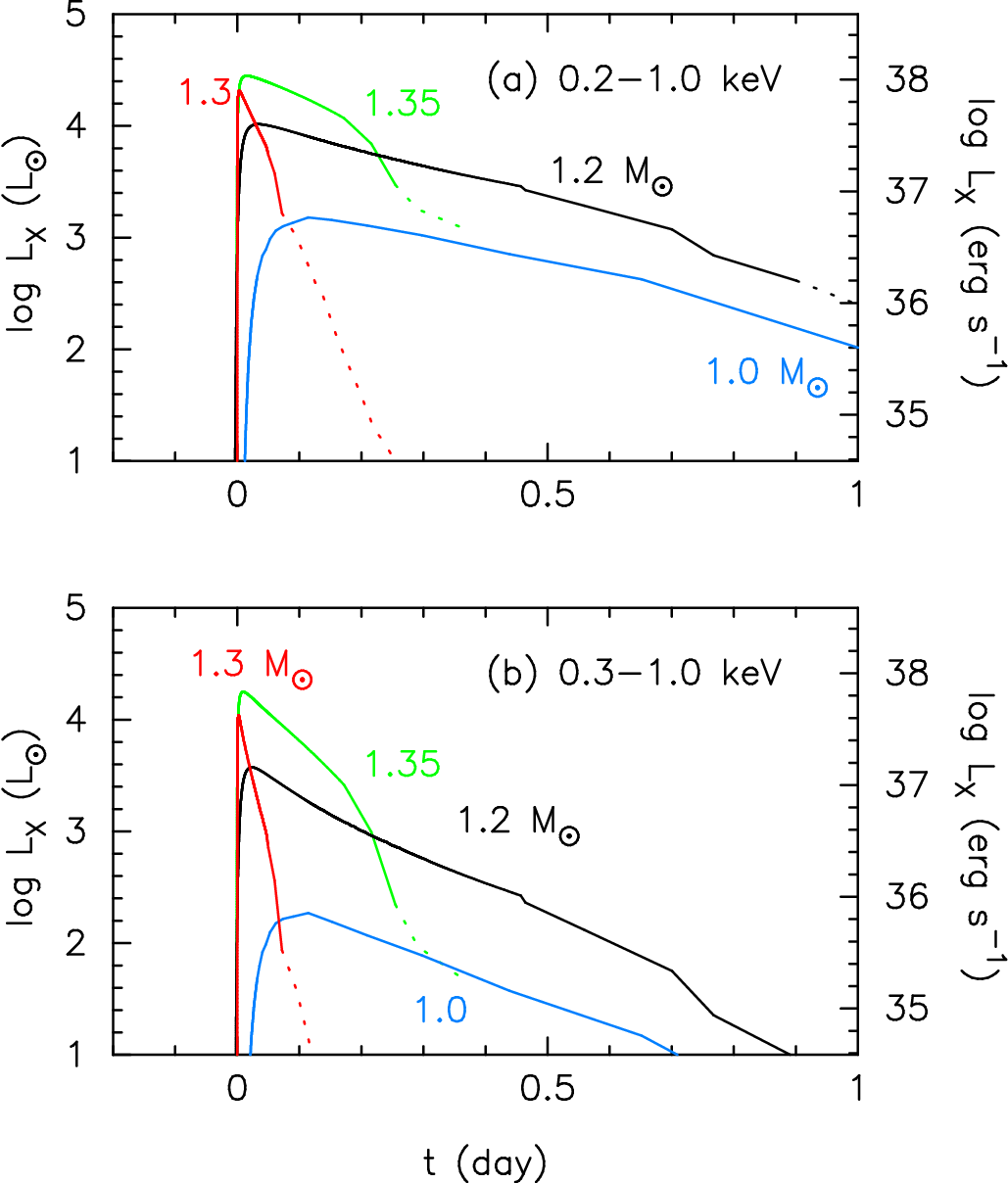}
\caption{
Close up views of X-ray light curves during X-ray flashes in  
(a) energy band of $0.2 - 1.0$ keV, and (b) $0.3 - 1.0$ keV. 
The dotted part indicates the wind phase, where the X-ray flux may not be
detected owing to self-absorption by the ejecta outside the photosphere.
The 1.3 $M_\odot$ WD model has the mass accretion rate of
$\dot M_{\rm acc}=2\times 10^{-9}~M_\odot$ yr$^{-1}$ and an envelope
which are mixed with carbon after hydrogen ignites. 
The other models, 1.35, 1.2, and 1.0 $M_\odot$ WDs are
for the same mass-accretion rate 
of $5\times 10^{-9}~M_\odot$ yr$^{-1}$ with the solar composition envelope.
The 1.0 $M_\odot$ WD model is taken from \citet{kat22sh},
and 1.35 $M_\odot$ WD model from \citet{kat22shapjl}. 
}\label{xrayflash}
\end{figure}

\subsection{X-ray flash}

Figure \ref{xrayflash} shows X-ray light curves during their X-ray flash 
phases for four different WD masses.
The 0.2-1.0 keV band corresponds to the SRG/eROSITA instrument, 
and 0.3-1.0 keV does to the Swift/XRT. 
Because most of photons are emitted below 1.0 keV, 
the X-ray luminosity hardly changes even   
if we adopt a higher upper limit, e.g., 10 keV.

The peak flux is higher for more massive WDs because of higher photospheric 
temperature and luminosity as shown in Figure \ref{hr}. 
The X-ray flush decays fastest in the C-rich 1.3 $M_\odot$ WD model 
followed by the solar-composition models in the order of the  
1.35, 1.2, and 1.0 $M_\odot$ WDs.

Figure \ref{xrayflash} demonstrates the dependence of
the X-ray flux on the WD mass as well as the energy band. 
Less massive WDs show lower fluxes especially in the $0.3-1.0$ keV band 
because a substantial part of the radiation is emitted below 0.2 keV. 
Thus, it is unlikely that X-ray flashes are detected in low mass WDs 
($M_{\rm WD} \lesssim 1.0 M_\odot$). 

The recent detection of an X-ray flash in the classical nova YZ Ret 
\citep{kon22wa} provides a rare opportunity to confirm  
theoretical models. 
The X-ray flash in YZ Ret was detected only once during the SRG/eROSITA 
all sky survey with four hours cadence, indicating that the X-ray flash
lasted only briefly ($<$ 8 hr). 
The blackbody temperature and luminosity is estimated to be  
$T_{\rm BB}= 3.27^{+0.11}_{-0.33} \times 10^5$ K and 
$L_{\rm ph}=(2.0 \pm 1.2) \times 10^{38}$ erg s$^{-1}$ \citep{kon22wa}, 
respectively. 
These properties suggest that the WD is very massive as examined 
by \citet{kat22shapjl} and \citet{kat22shc}. 
\citet{hac23yzret} presented a
theoretical model of YZ Ret, in which a 1.33 $M_\odot$ WD reproduces
the multiwavelength light curves, 
from early gamma-ray emission to later supersoft X-ray phase,
as well as optical light curves.

From the X-ray spectrum analysis of YZ Ret, 
\citet{kon22wa} found no major intrinsic absorption during 
the X-ray flash.
Thus, \citet{kat22shapjl} concluded that \\
(1) no dense matter exists around the WD photosphere,\\
(2) no indication of a shock wave, and\\
(3) the hydrogen-rich envelope is almost hydrostatic.\\
Note that all of these are consistent with our theoretical model. 
Optically-thick winds do not yet emerge from the photosphere
at the epoch of X-ray bright (X-ray flash) phase
and the envelope is almost hydrostatic.  


 
Our X-ray light curves in Figure \ref{xrayflash} suggest that 
we can expect bright X-ray flash for massive WDs ($> 1.0~M_\odot$) before 
the optical brightening, except the WD is deeply embedded in a dense 
gaseous matter likely in symbiotic stars.

\subsection{Internal structure of the envelope}
\label{sec_structure}

\begin{figure}
  \includegraphics[width=8cm]{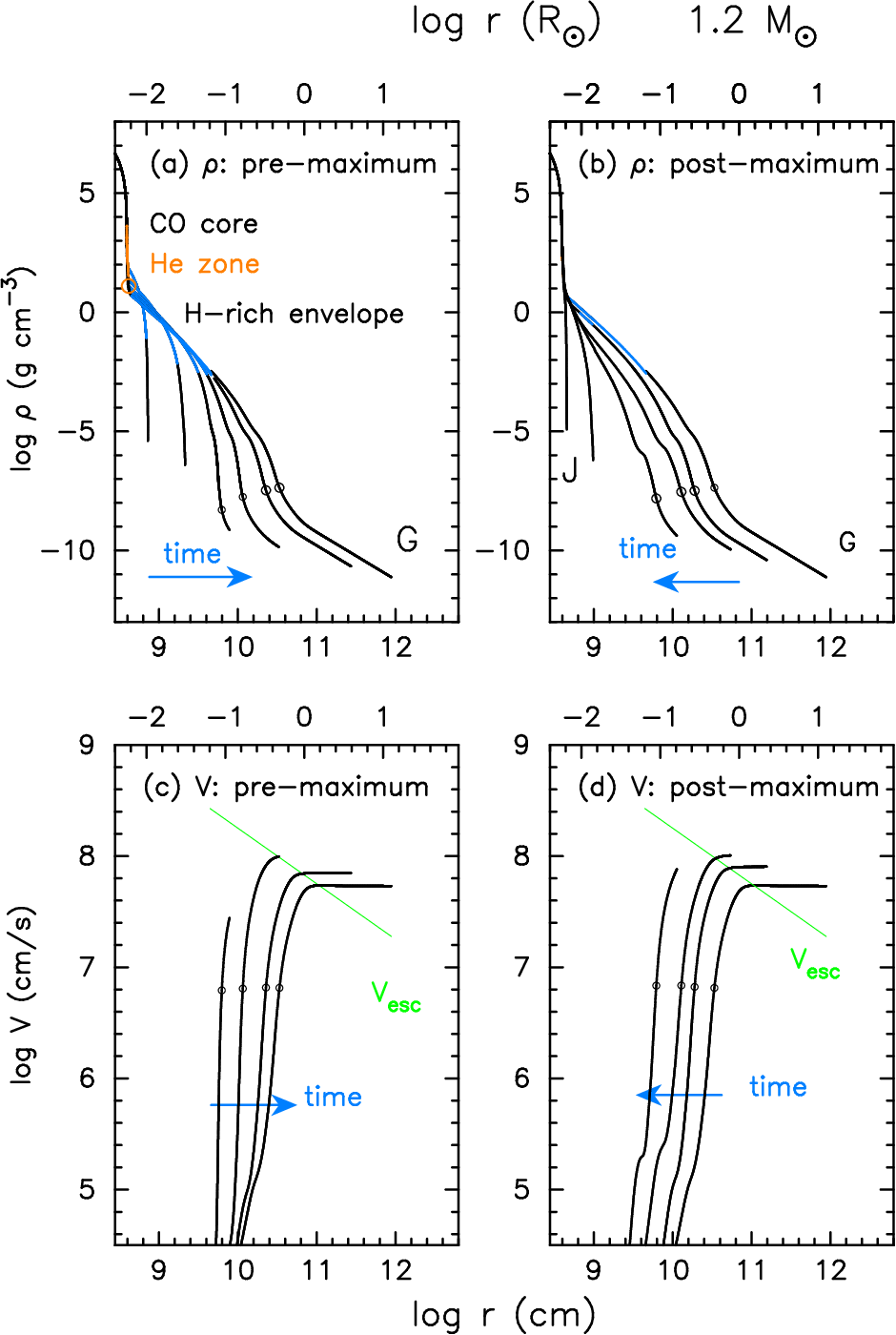}
\caption{Temporal structure changes in the  $1.2~M_\odot$ WD model. 
The density profiles toward optical maximum (a) and after maximum (b).
The velocity profiles toward optical maximum (c) and after maximum (d).
Here, $r$ is the radius from the center of the WD.  
The outermost point of each line corresponds to the photosphere. 
The blue lines indicate the convective region. 
The critical points of each wind solution are indicated by
the small open black circles.  The green straight line shows the escape
velocity $v_{\rm esc}=\sqrt {2 G M_{\rm WD}/r}$.
In panel (a), the region of CO core is denoted by the black line while
the helium layer is by the orange line.  The open orange circle represents
the place of maximum energy generation rate by nuclear burning
$\epsilon_{\rm nuc}^{\rm max}$.
}\label{rhov.m12}
\end{figure}

\begin{figure}
  \includegraphics[width=8cm]{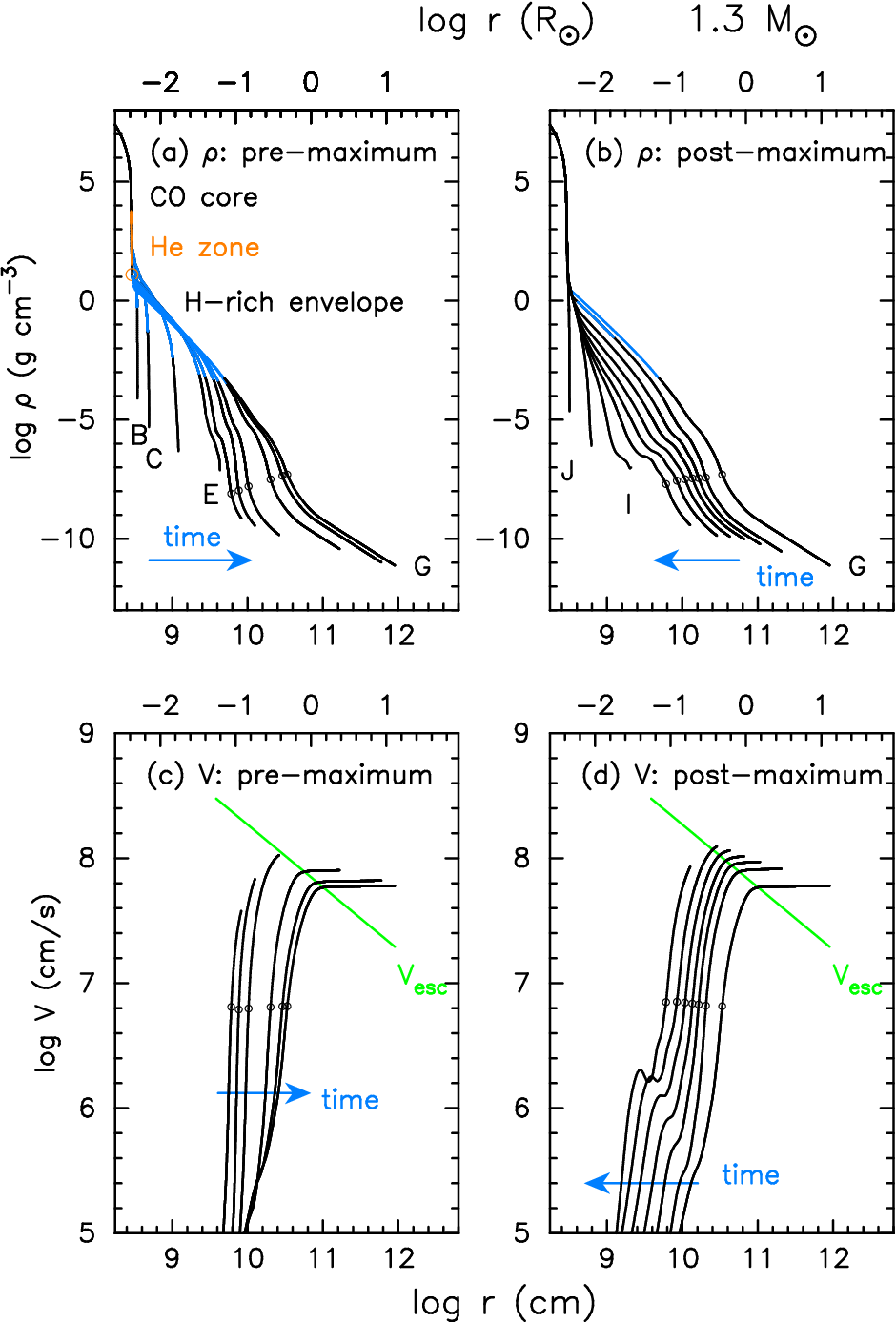}
\caption{Same as those in Figure \ref{rhov.m12}, but for the 1.3 $M_\odot$
WD model. 
}\label{rhov.m13}
\end{figure}

Figures \ref{rhov.m12} and \ref{rhov.m13} show the temporal changes of
the density and velocity profiles in the 1.2 and 1.3 $M_\odot$ WDs,
respectively, from an early stage to optical maximum (epoch G), and 
from optical maximum (G) to the end of the outburst (J).  
Convection occurs above the hydrogen burning zone (blue line region) 
from the beginning of thermonuclear runaway to the extended 
stages of 
nova envelope, and then disappears shortly after epoch G. 

Optically thick winds are accelerated to emerge from the photosphere
when the photospheric temperature decreases and approaches the 
prominent Fe opacity peak at $\log T$ (K) $\sim$ 5.3.  
The velocity quickly increases around the critical point of the wind
\citep{bon52, kat94h}, where the velocity reaches the sound velocity. 
We did not include convective energy transport above the critical point, 
because the wind velocity is supersonic and therefore 
convective eddies cannot turn back.

In the pre-maximum phase [Figures \ref{rhov.m12}(c) and \ref{rhov.m13}(c)] 
the acceleration region (around the opacity peak: $\log T$ (K) $\sim 5.2$) 
moves outward. As the envelope expands the radius at the critical point
increase, the density of which also increases. 
Thus, the wind mass loss rate, 
$\dot M_{\rm wind}=4 \pi r_{\rm cr}^2 \rho_{\rm cr} v_{\rm cr}$, 
increases with time.  The velocity 
reaches the terminal velocity deep inside the photosphere, 
because the winds are hardly accelerated outside the opacity peak. 

In this way, the wind mass-loss rate increases,  
but the velocity at the photosphere decreases in the pre-maximum phase.   
On the contrary, in the post-maximum phase, 
the wind mass-loss rate decreases 
whereas the velocity at the photosphere increases with time. 
These trends are common among the three, i.e., 1.0 $M_\odot$,
1.2 $M_\odot$, and 1.3 $M_\odot$ WD models.

\citet{hac22shock} proposed a strong shock formation in the nova ejecta 
(so-called internal shock)
based on the 1.0 $M_\odot$ model \citep{kat22sh}.  
Assuming that the ejecta motion is ballistic above the photosphere,
they proposed a mechanism of strong shock formation outside the photosphere.
In the pre-maximum phase, the later ejected matter has a smaller velocity 
so the ejected matter/gas expands.  In the post-maximum phase, however,
later ejected matter has a larger velocity that will catch up the earlier
ejected matter, which causes a strong compression and then a shock.  
They predicted that a strong shock wave forms only after optical maximum.  

This shock formation naturally explains hard X-ray emission  
\citep{fri87, muk01i} and absorption/emission line systems such as principal
and diffuse-enhanced systems \citep{mcl42}, as discussed by \citet{hac22shock}
and \citet{hac23yzret}. 
Our 1.2 and 1.3 $M_\odot$ WD models provide theoretical supports
to the internal shocks for various fast novae. 

Figure \ref{rhov.m13} also shows similar profiles, but for the 1.3 $M_\odot$
WD model. 
The small hump in the velocity profile deep inside the 
critical point is caused by a small opacity peak 
owing to C and O at $\log T$ (K) $\approx 6.1$. 
Such a velocity hump does not appear in the 1.0 and 1.2 $M_\odot$ WD models. 
Note that carbon is not enhanced in the envelopes of 1.2 and 1.0 $M_\odot$
WD models.

\section{Energy budget} \label{sec_energy}

\begin{figure}
  \includegraphics[width=8cm]{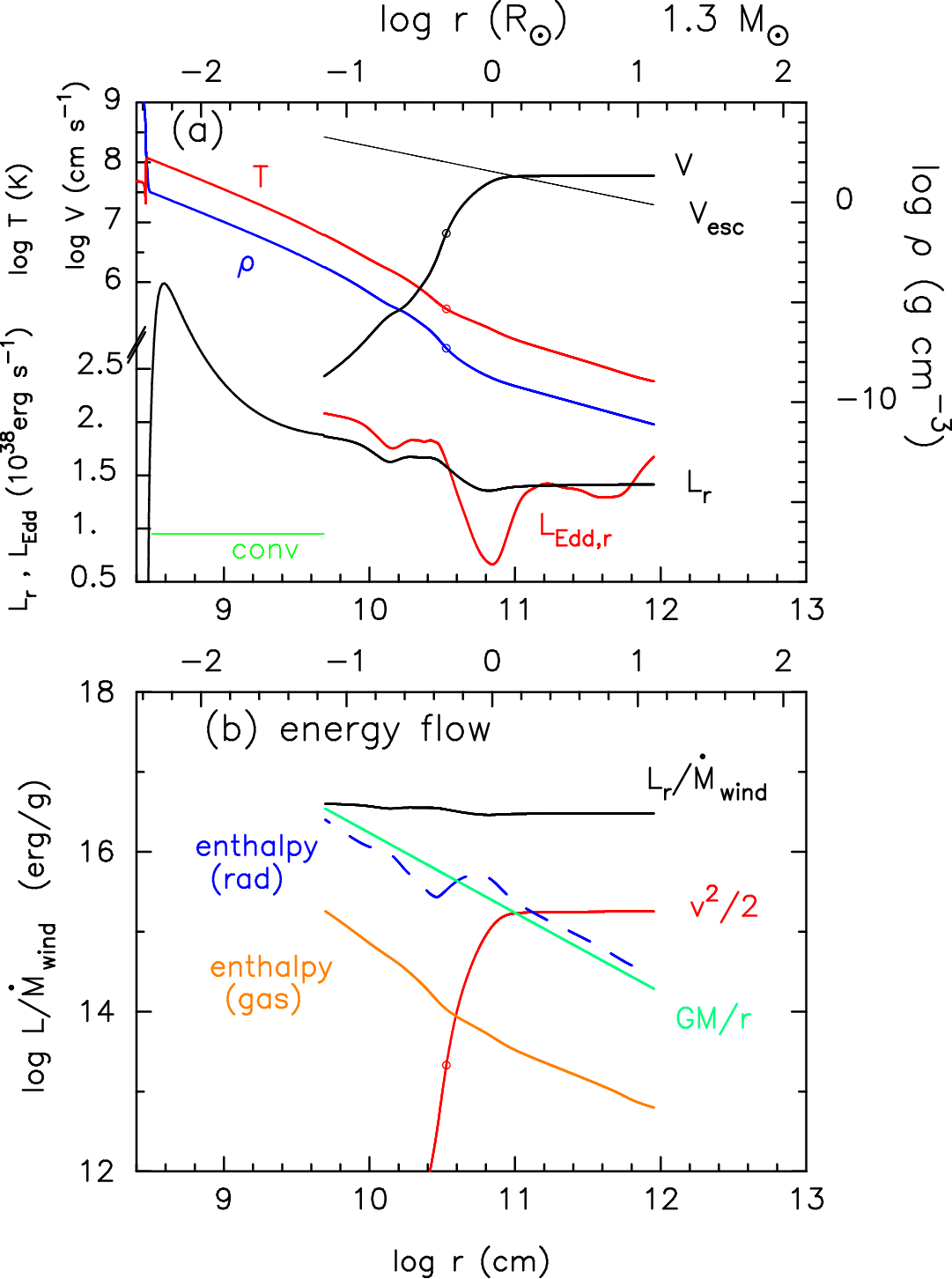}
\caption{Internal structure of the envelope at the maximum expansion of the
photosphere (at epoch G) for the 1.3 $M_\odot$ WD model.
(a) The distribution of the temperature (red line), density (blue line), 
velocity (black), local luminosity $L_{r}$, and local Eddington 
luminosity. The outermost point of each line corresponds to the photosphere.
The critical point of solar-wind type solution \citep{kat94h} is 
indicated by open circle. 
(b) The distribution of the diffusive radiative luminosity
per unit mass, $L_r/\dot M_{\rm wind}$, and energy advected
by winds: gravitational energy $GM/r$,
radiation enthalpy (rad), gas enthalpy (gas),
and kinetic energy $v^2/2$.
}\label{struc.max}
\end{figure}

Figure \ref{struc.max}(a) shows the internal structure of 
the envelope at epoch G of the 1.3 $M_\odot$ WD model.  The 1.2 $M_\odot$
WD has a similar structure to that of the 1.3 $M_\odot$ WD, and 
thus, we do not discuss it here. 
The local luminosity $L_{r}$ is defined by the sum of radiative and 
convective luminosities, where $r$ denotes the radius from the center of
the WD.  The local luminosity $L_r$ steeply increases
in the nuclear burning region,
and then turns to decrease because a substantial part of $L_r$ is absorbed
above the zone. 
The convective region is indicated by the horizontal green line in
Figure \ref{struc.max}(a).  The envelope is convective only 
in the innermost part, where $L_{r}$ is large. 

We plot in Figure \ref{struc.max}(a) the local Eddington luminosity, 
\begin{equation}
L_{{\rm Edd},r} \equiv {4\pi cG{M_{\rm WD}} \over\kappa_r},
\label{equation_Edd,r}
\end{equation}
where $\kappa_r$ is the local opacity which is a function of the density
and temperature for a given chemical composition.
A small dip of $L_{{\rm Edd}, r}$ at $\log$ $r$ (cm) $\sim$10.1
in Figure \ref{struc.max}(a)
corresponds to a small peak in the opacity contributed by ionized C and O 
($\log T$ (K) $\sim 6.1$--6.3)
and a large dip of $L_{{\rm Edd}, r}$
at $\log$ $r$ (cm) $\sim 10.85$ is caused by
a large Fe peak at $\log$ $T$ (K) $\sim 5.2$. 
The wind is accelerated where the opacity increases outward
and $L_{r}$ exceeds the local Eddington luminosity $L_{\rm Edd,r}$
as shown in Figure \ref{struc.max}(a).

During the wind mass-loss phase, the ejecta carry energy away 
from the WD. As introduced in \citet{kat22sh}, we integrated each term of 
the energy conservation equation, i.e.,
\begin{equation}
L_r-\dot M_{\rm wind} ({v^2\over 2}+\omega_{\rm rad}+
\omega_{\rm gas}-{G M\over r}) = \Lambda,
\label{equation_Lr}
\end{equation}
and obtained each energy budget \citep{kat94h}.
The second term in the left-hand-side of equation (\ref{equation_Lr})
is the energy carried with the moving matter, that includes 
the kinetic energy, enthalpy of radiation $\omega_{\rm rad}$
(photon energy trapped in the moving matter),
enthalpy of gas $\omega_{\rm gas}$, and gravitational energy.

Figure \ref{struc.max}(b) depicts the distribution of each energy flux
divided by $\dot{M}_{\rm wind}$. 
The enthalpy of radiation is decreasing outward
to compensate the gravitational energy, and 
as a result, the diffusive luminosity $L_r/\dot{M}_{\rm wind}$ becomes a
dominant term near the photosphere.  These properties are essentially
the same as those of the 1.0 $M_\odot$ WD model in \citet{kat22sh}. 

The diffusive luminosity $L_r$ substantially decreases outward at $\log$~$r$
(cm) $\sim 10.7$, where the opacity increases outward and therefore 
the local Eddington luminosity decreases outward. 
The decreased radiative energy flux is consumed for wind mass loss, 
supplying the gravitational energy of the envelope and 
kinetic energy of the winds.


\begin{longtable}{*{8}{c}}
 \caption{Energy budget}\label{table.wind}
  \hline              
 &    subject  &$E_{\rm nuc}$&$M_{\rm ej}$ & $E_{\rm rad}$&$E_{\rm G}$&$E_{\rm kin}$  \\ 
 &  unit  &(erg)&($M_\odot$) &(erg) &(erg) & (erg) \\
\endfirsthead
  \hline
  Name & Value1 & Value2 & Value3 \\
\endhead
  \hline
\endfoot
  \hline
\endlastfoot
  \hline
$1.3~M_\odot$&  &$4.0\times 10^{45}$&$1.8\times 10^{-6}$&$1.9\times 10^{45}$ &$2.1\times 10^{45}$& $1.2\times 10^{43}$ \\
\% &  & 100 & --   & 47 & 52 & 0.29\\ 
  \hline
$1.2~M_\odot$&  &$1.4\times 10^{46}$&$6.7\times 10^{-6}$&$8.3\times 10^{45}$ &$5.2\times 10^{45}$& $3.8\times 10^{43}$ \\
\% &  & 100 & --   & 61 & 39 & 0.28\\ 
  \hline
$1.0~M_\odot$  & &$3.2\times 10^{46}$ &$3.0\times 10^{-5}$ &$2.1\times 10^{46}$ & $1.1\times 10^{46}$  &$7.3 \times 10^{43}$ \\
\% & &  100 &-- & 64 &35 &0.23 \\
\end{longtable}

Following \citet{kat22sh}, we estimate how much amount of nuclear energy
is consumed to drive the wind mass-loss.  We have calculated the total nuclear
energy generated during one cycle of a nova outburst to be 
$E_{\rm nuc}=\int L_{\rm nuc} dt$, energy emitted as radiation from
the photosphere $E_{\rm rad}=\int L_{\rm ph} dt$, and the kinetic 
and gravitational energy carried out in the wind 
$E_{\rm kin}=\int {1 \over 2} \dot M_{\rm wind}v^2_{\rm ph} dt$, 
and $E_{\rm G}=\int GM_{\rm WD}\dot M_{\rm wind}/R_0 dt$, 
respectively, where $v_{\rm ph}$ is the wind velocity at the photosphere and
$R_0$ the radius of nuclear burning region, $\log (R_0/R_\odot)=-2.229$
for 1.2 $M_\odot$ and $\log (R_0/R_\odot)=-2.365$ for 1.3 $M_\odot$.  

Table \ref{table.wind} summarizes the energy budget for our models.  
The total nuclear energy $E_{\rm nuc}$ is smaller for a more
massive WD, because the ignition mass is much smaller.
Correspondingly each allocated energies are smaller for a more massive WD.  

Most of the energy is used in the photospheric emission (blackbody emission) 
and gravitational energy of the wind.
Even in the strong shell flash of the 1.3 $M_\odot$ WD,
the allocated kinetic energy 
is as small as less than 1\% of the total nuclear energy.


\section{Discussion}\label{sec_discussion}

\subsection{Pre-maximum evolution of fast novae}

Pre-maximum evolution of a nova outburst has been poorly understood.
For example, the timescale from the outburst to optical maximum, 
$t_{\rm peak}$, is not well constrained for many novae, contrary to
the decay timescales such as $t_2$ and $t_3$, which are the days by
2 and 3 mag decay from the optical ($V$) maximum.  
Table \ref{table_time} shows the evolution time from the ignition to 
the maximum expansion (stage G) to be $t_{\rm peak}=26$ day
in model M10 (1.0 $M_\odot$) 
and 11 day in M12 (1.2 $M_\odot$). 
Although M135 model has stopped before reaching stage G 
because of numerical difficulties, we estimate the 
time to stage G to be $t_{\rm peak} \sim 7$ day.
In these models, the WDs accrete matter of the same chemical composition
($X=0.7$, $Y=0.28$, and $Z=0.02$) at the same mass-accretion rate
of $\dot M_{\rm acc}=5\times 10^{-9}~M_\odot$ yr$^{-1}$. 
For the carbon enhanced model M13C10
($\dot M_{\rm acc}=2\times 10^{-9}~M_\odot$), 
the time from the ignition to stage G is as short as 2.5 day 
as expected from the short total duration of the outburst.   
We can see a tendency that the time to optical maximum is longer in 
a less massive WD (or shorter in a more massive WD). 

It is difficult to observationally detect when the outburst begins. 
Many novae have been discovered near their optical peaks (stage G).  
Only the exception is the classical nova YZ Ret \citep{kon22wa} in which 
an X-ray flash was detected for the first time, after 
unsuccessful attempts for other novae  
\citep{mor16,kat16xflash}. 
The X-ray flash begins immediately after the thermonuclear runaway 
(see Table \ref{table_time}), 
and lasts only less than one day (Figure \ref{xrayflash}). 

In YZ Ret, the time of optical peak is not well constrained
but less than four days after the X-ray flash.
The WD mass is estimated as massive as 
$\gtsim 1.3 ~M_\odot$ from multiwavelength light curve fitting 
\citep{hac23yzret}. This short timescale ($t_{\rm peak}\lesssim 4$ day)
is consistent with the decreasing tendency of $t_{\rm peak}$ 
for a more massive WD in Table \ref{table_time}.  

To formulate the tendency of $t_{\rm peak}$, we define the timescale
$\tau_{\rm exp}$ of expansion of the envelope at point B
(maximum nuclear burning rate) by
\begin{equation}
\tau_{\rm exp} L_{\rm nuc}^{\rm max} = {{G M_{\rm WD}} \over R_{\rm WD}} 
M_{\rm env},
\label{def_expansion_timescale}
\end{equation}
where $M_{\rm env}$ is the envelope mass at ignition, i.e., 
$M_{\rm env}= M_{\rm ig}$, and $R_{\rm WD}$ is the WD radius and
we adopt $R_{\rm WD}= R_0$ at the radius of the bottom nuclear burning region.
This relation presents an approximate balance between the nuclear energy 
relased during the expansion and the work needed to lift-up the graviational energy 
of the envelope accumulated before ignition. 

\begin{figure}
  \includegraphics[width=8cm]{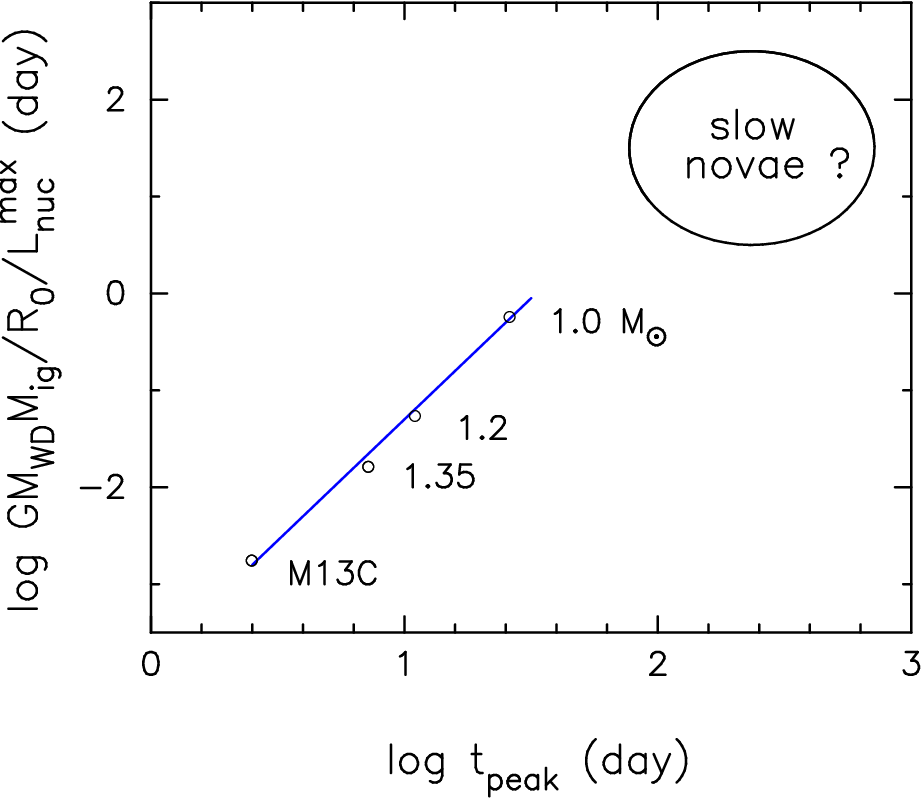}
\caption{
The expansion timescale ($\tau_{\rm exp}$) versus time to the optical peak
($t_{\rm peak}$) for fast novae, where $\tau_{\rm exp}\equiv
G M_{\rm WD} M_{\rm ig}/ R_0 / L_{\rm nuc}^{\rm max}$ from Equation
(\ref{def_expansion_timescale}).  The blue line has a slope of
$t_{\rm peak}\propto (\tau_{\rm exp})^{0.4}$.
}\label{expansion_versus_peak_time}
\end{figure}

We plot the $\log t_{\rm peak}$-$\log \tau_{\rm exp}$ diagram in Figure
\ref{expansion_versus_peak_time}.
Our four models, M10, M12, M135, and M13C, are located almost on a
straight blue line in the diagram.  Thus, our expansion timescale
$\tau_{\rm exp}$ defined by Equation (\ref{def_expansion_timescale})
correctly indicates the time to the optical ($V$) peak $t_{\rm peak}$,
independently of the envelope chemical composition, mass-accretion rate,
or WD mass.  This means that we predict $t_{\rm peak}$ if we calculate 
$L_{\rm nuc}^{\rm max}$, $R_{\rm WD}= R_0$, and $M_{\rm env}= M_{\rm ig}$
at the very early phase of nova outburst. From the 
straight line in Figure \ref{expansion_versus_peak_time} we obtain 
the relation; 
\begin{equation}
t_{\rm peak} \approx 33 \left({{\tau_{\rm exp}} \over {{\rm day}}}
 \right)^{0.4} {\rm ~day}.
\label{peak_time_vs_expansion_time}
\end{equation}
These $L_{\rm nuc}^{\rm max}$, $R_0$, and $M_{\rm ig}$ 
are relatively easily obtained
without computing very expanded stages such as stage G and,
as a result, $\tau_{\rm exp}$ is also easily obtained with 
our Henyey code because no winds are accelerated yet at point B.

\subsection{Pre-maximum evolution of slow novae}
Pre-maximum phases have been partly observed in some slow novae.  
In the extremely slow nova PU Vul, the $m_{\rm pv}$ brightened 
from 11.2 to 9.55 mag during 127 days \citep{wen79}.
The WD mass in PU Vul is estimated, from the UV 1455\AA~ light curve fitting in 
the optical decay phase, to be about 0.6  $M_\odot$ \citep{kat11}. 
In the slow nova CN Cha, it takes 300 days to rise from $V=16$ to $V=10$
\citep{lan20}.
The WD mass is estimated be about 0.57- 0.6 $M_\odot$ \citep{kat23}.  
In the slow novae V723 Cas, it took 24 days to rise from unfiltered
CCD magnitude 12.2 to 9.2 \citep{ohshi95}.
The WD mass possibly as small as 0.6-0.7 $M_\odot$ \citep{kat11drag}.   
For nova Vel 2022 (Gaia22alz), the $g$ magnitudes rose from 17.8 to
13 in 100 days \citep{ayd23}. 
The WD mass is not known but the light curve resembles 
to the slow nova V723 Cas. 
We have no information on the outburst (ignition) time of these novae, but 
thermonuclear runaway should occur before the optical brightening.  
Thus, the duration between the onset of the outburst and the optical peak 
should be longer than these 23 - 300 days.  It is probably 
a few hundred days or more for less massive WD ($\sim 0.6~M_\odot$).   
Such a long duration is consistent with the tendency in the 
lower mass limit in Table \ref{table_time} and in Figure
\ref{expansion_versus_peak_time}. 


\section{Concluding remarks}\label{conclusions}

Novae show rich variety in their timescales, peak luminosities, 
speed classes, light curve shapes, spectral developments,
ejecta properties etc.  Such variety is a consequence of various WD masses, 
mass accretion rates, and binary parameters.  To reveal the overall picture
of nova phenomena, multiwavelength observations,  
along different stages of a nova outburst, play essential roles. 
Very early rising phase of a nova is an unexplored field,
which has just been opened by the detection of an X-ray flash in YZ Ret. 
Our full-cycle shell flash models provide information of such 
a very early phase of a nova outburst which could be
helpful in the massive survey era. 
  



\end{document}